\documentclass[journal=jpccck,manuscript=article]{achemso}
\usepackage{chemformula} 
\usepackage[T1]{fontenc} 
\usepackage{amsmath}
\usepackage[labelfont=bf]{caption}
\usepackage{lineno}
\usepackage{lipsum}

\author{Sayan Routh}
\author{Setti Thirupathaiah}
\affiliation{Department of Condensed Matter Physics and Material Sciences, S N Bose National Centre for Basic Sciences, Kolkata, West Bengal-700106, India}
\email{setti@bose.res.in}

\title{Observation of Exchange Bias in Antiferromagnetic Cr$_{0.79}$Se due to Coexistence of Itinerant Weak Ferromagnetism at Low-temperature}

\begin{document}

\setlength\linenumbersep{1.5cm}
\renewcommand\linenumberfont{\normalfont\bfseries\small}

\begin{abstract}
We report on the structural, electrical transport, and magnetic properties of antiferromagnetic transition-metal monochalcogenide Cr$_{0.79}$Se. Different from the existing off-stoichiometric compositions, Cr$_{0.79}$Se is found to be synthesised into the same NiAs-type hexagonal crystal structure of CrSe. Resistivity data suggest Cr$_{0.79}$Se to be a Fermi-liquid-type metal at low temperatures, while at intermediate temperatures the resistivity depends sublinearly on the temperature.  Eventually,  at the elevated temperatures the rate of change of resistivity rapidly decreases with increasing temperature. Magnetic measurements suggest a transition from paramagnetic phase to an antiferromagnetic phase at a N$\acute{e}$el temperature of 225 K. Further reduction of the sample temperature results into coexistance of weak ferromagnetism  along with the antiferromagnetic phase below 100 K. As a result, below 100 K, we identify significant exchange bias due to the interaction between the ferro- and antiferromagnetic phases. In addition, from the temperature dependent X-ray diffraction measurements we observe that the NiAs-type structure is stable up to as high as 600$^o$C.
\end{abstract}

\maketitle

Keywords: Magnetism, Exchange Bias, Transition-metal Monochalogenides, Metal-Insulator Transition

\section{Introduction}
Design and synthesis of materials with a strong magnetic exchange bias (EB) property has been one of the intense research activities from the past several decades~\cite{Meiklejohn1957} and till to the present days~\cite{Wang2011, Nayak2015, Saha2019, Tian2021} due to their potential applications in spintronic devices~\cite{Parkin2004, Hirohata2014}. There exists several studies on designing the multilayered and core-shell structures to generate an effective large exchange bias at the interface of a ferromagnetic (FM) and an antiferromagnetic (AFM) layer~\cite{Inderhees2008, Lage2012, Lavorato2017, Perzanowski2017, Song2017}. Several bulk materials too have been synthesised, showing large exchange bias~\cite{Wisniewski2017, Belik2013, Fertman2020}. But most of the bulk materials are in the form of nanocomposites or with a complicated crystal structure of the doped ternary compounds. In this paper, we will discuss the exchange bias in a transition-metal monochalcogenide having a simplest crystal structure.

\begin{figure}[ht]
\centering
  \includegraphics[width=0.45\textwidth]{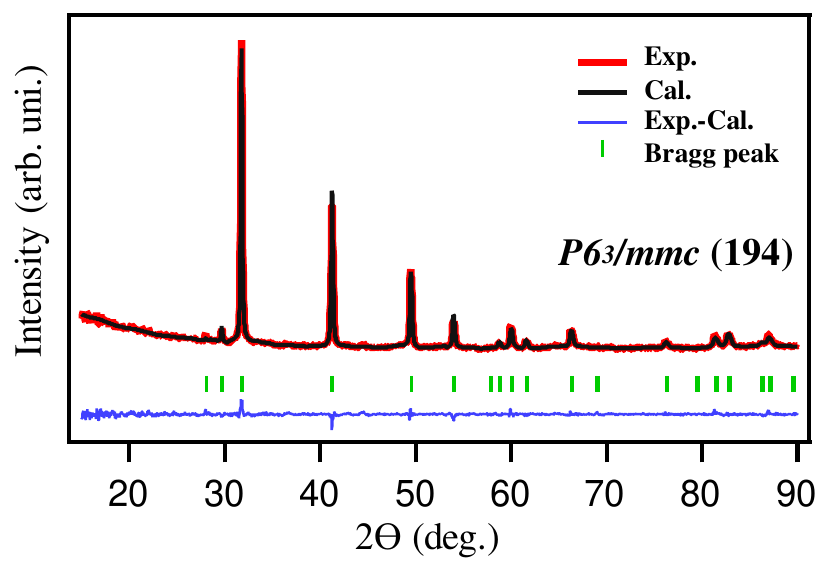}
  \caption{Powder X-ray diffraction pattern and reitveld refinement of Cr$_{0.79}$Se, confirming the NiAs-type crystal structure with  space group of P6$_3$/mmc (194).}
  \label{1}
\end{figure}

Transition-metal monochalcogenides with the chemical formula of MX (M= Fe, Cr; X=S, Se, Te)  are very versatile materials due to their diverse structural, electronic, and magnetic properties. For instance, Fe$_x$Se is a non-magnetic high temperature superconductor with a T$_c$ of 8 K,  having a tetragonal crystal structure for x$\geq$1~\cite{Hsu2008}, while it is an antiferromagnetic metal having hexagonal crystal structure for x$<$1~\cite{Li2016}. Whereas, FeTe is always a tetragonal antiferromagnetic system with a stripe order~\cite{Maheshwari2015, Haenke2017}. Further, FeS is found to be a non-magnetic superconductor with a tetragonal crystal structure~\cite{Zhang2017}.  On the other hand, similar to Fe$_x$X, Cr$_x$X (X=S, Se, Te) systems too are very diverse in their structural,  electronic, and magnetic properties~\cite{Chen2019, Yang2019, Li2019, Sun2020, Coughlin2020, Huang2021}. For instance, Cr$_x$Te is a ferromagnetic half-metal and can exist in any of zinc-blend (ZB)~\cite{Sanyal2003}, rock-salt (RS)~\cite{Liu2010}, or NiAs~\cite{Lotgering1957} crystal structure type. Whereas,  Cr$_{x}$S~\cite{Kamigaichi1960} and Cr$_{x}$Se~\cite{Corliss1961} are mostly known for their antiferromagnetic nature having the NiAs-type crystal structure. Some reports suggested Cr$_{x}$Se to be even a spin-glass type magnetic system~\cite{Li2006}  and Cr$_{x}$S to be a ferrimagnetic metal~\cite{Konno1988}.

\begin{figure*}[ht]
\centering
  \includegraphics[width=0.9\textwidth]{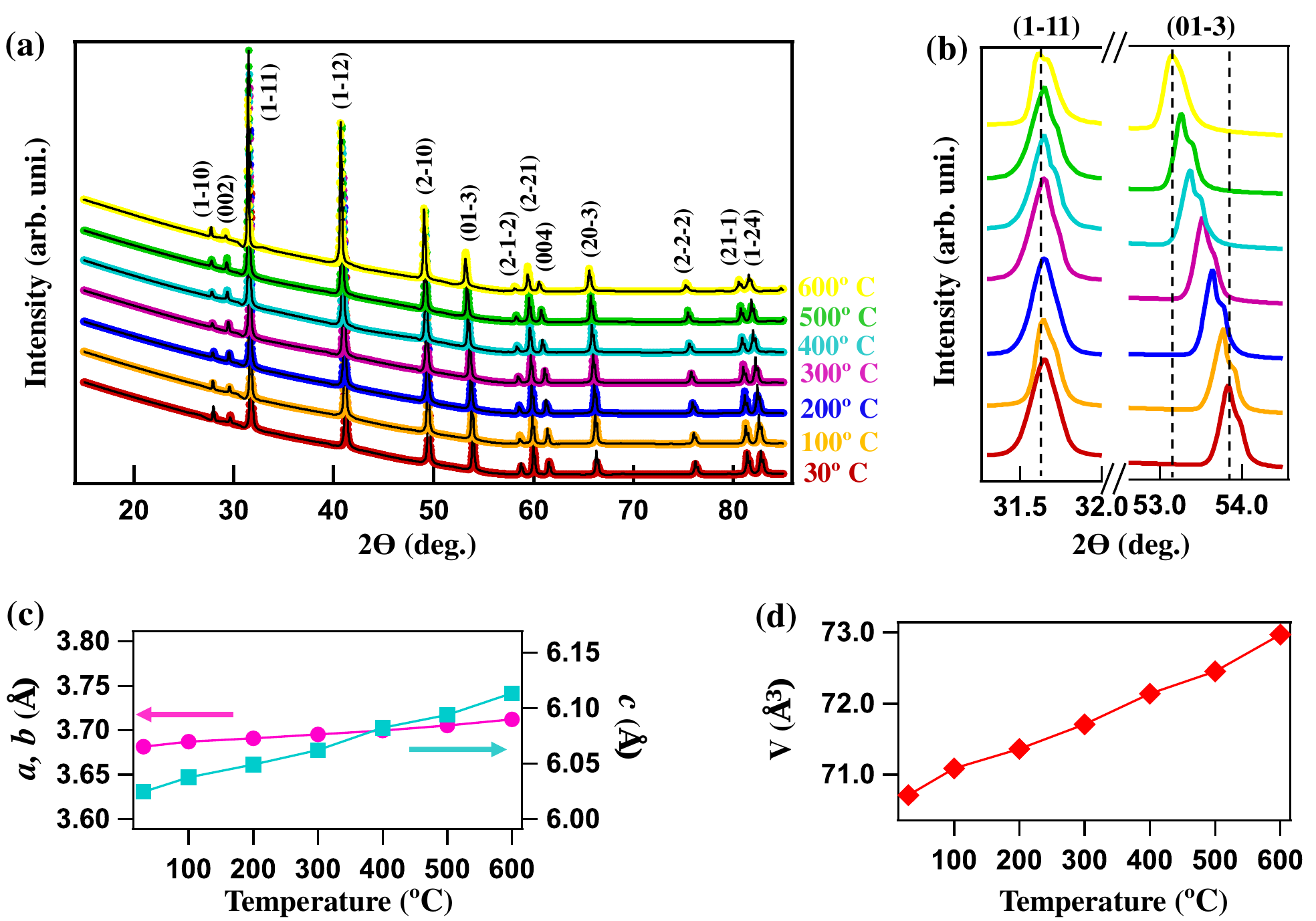}
  \caption{(a)Temperature dependent powder X-ray diffraction pattern of Cr$_{0.79}$Se overlapped with reitveld refinement. (b) Enlarged XRD patterns for the reflections  of (1$\bar{1}$1) and (01$\bar{3}$). (c) Plot of lattice constants $a$, $b$, and $c$ as a function of temperature. (d) Plot of cell volume ($V$) as a function of temperature.}
  \label{2}
\end{figure*}

In this paper, we report a comprehensive study on the structural, electrical transport, and magnetic properties of Cr$_{0.79}$Se in the polycrystal form. Till date not many experimental studies are available on these system, despite being a non-collinear AFM metal~\cite{Corliss1961}. Recently, it was suggested that the antiferromagnetic metals with non-collinear spin texture are promising candidates for the anomalous Hall effect,  induced by the berry curvature~\cite{Gan2016, Yan2017, Li2020}. With this motivation, we reinvestigated the structural, electrical, magnetic properties of this system. Our x-ray diffraction (XRD) studies demonstrate that Cr$_{0.79}$Se has the NiAs-type structure. At higher temperatures we noticed shift in certain XRD peak positions, leading to change in lattice parameters with temperature. In addition, from the XRD measurements we observe that the NiAs-type structure is stable up to as high as 600$^o$C of sample temperature. Electrical resistivity studies show Fermi-liquid like metallic behaviour at low temperatures ($<41$ K), and in the intermediate temperatures (41-200 K) the resistivity changes sublinearly with temperature.  Further, at the elevated temperatures ($>200$ K) the rate of change of resistivity rapidly decreases with temperature. Magnetic properties studies suggest a transition from paramagnetic phase to an antiferromagnetic phase at a N$\acute{e}$el temperature of 225 K. Further, below 100 K, a weak ferromagnetism is found which is coexisting with antiferromagnetism.


\begin{figure}[ht]
\centering
  \includegraphics[width=0.45\textwidth]{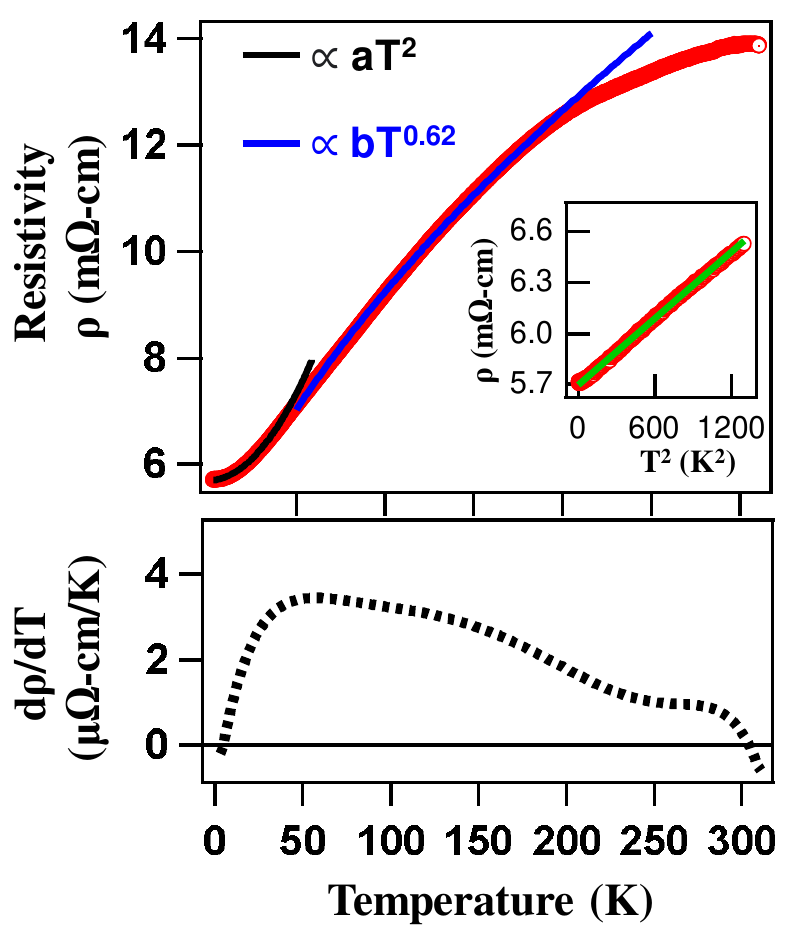}
  \caption{Temperature dependent electrical resistivity of Cr$_{0.79}$Se. Black solid curve represents T$^2$ law fitting up to 41K and the blue solid curve represents sublinear fitting between 41 and 200 K. Inset shows the plot of $\rho$ $vs.$ $T^2$. Green line in the inset is a linear fit to the data. Bottom image represents d$\rho$/dT $vs.$ T}
  \label{3}
\end{figure}


\section{Results}

Figure~\ref{1}(a) shows rietveld refinement on the XRD data of Cr$_{0.79}$Se measured at the room temperature (RT). It is evident from the XRD data that Cr$_{0.79}$Se crystallizes into the NiAs-type crystal structure with a hexagonal space group of P6$_3$/mmc(194). The estimated lattice parameters from the rietveld refinement are found to be $a$=$b$=3.6811(3) Å and $c$=6.0198(6) Å.  No additional impurity peaks have been noticed from the XRD data, demonstrating high phase purity of the sample. Further, we have performed XRD measurements as a function of temperature starting from RT  to 600$^o$C, as shown in Figure~\ref{2}(a).  From the temperature dependent XRD data we noticed that the peak positions are relatively shifted with the temperature. To demonstrate the peak shift, in Figure~\ref{2}(b), we fixed peak position of the reflection (1$\bar{1}$1) to notice a significant shift in peak position of the reflection (01$\bar{3}$) . In order to elucidate the structural changes with the temperature, we performed rietveld refinement for the XRD data at every measured temperature. The obtained lattice parameters are plotted as a function of temperature as shown in Figure~\ref{2}(c). We identify that the lattice parameter $a(b)$ is almost constant,  changing from 3.681 Å to 3.712 Å, while the lattice parameter $c$ substantially increases from 6.024 Å to 6.113 Å in going from RT to 600$^o$C.  Consequently, the unit cell volume also increases with the temperature as shown in Figure~\ref{2}(d).



\begin{figure}[hb]
\centering
  \includegraphics[width=0.45\textwidth]{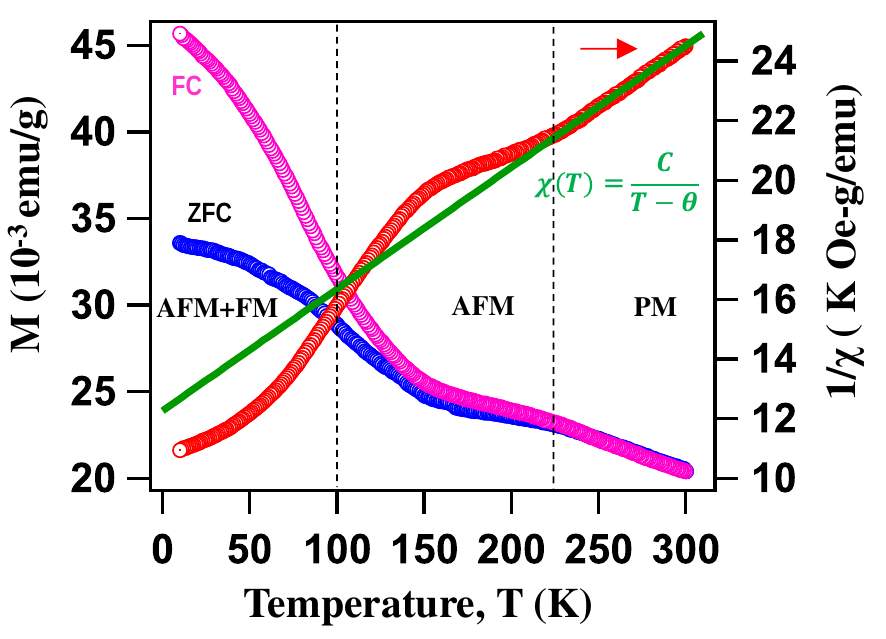}
  \caption{Magnetization as a function of temperature plotted for zero-field-cooled (ZFC) and field-cooled (FC) modes. Susceptibility as a function of temperature is plotted for the FC mode. Green curve is a susceptibility fitting with the Curie-Weiss law.}
  \label{4}
\end{figure}

\begin{figure}[ht]
\centering
  \includegraphics[width=0.45\textwidth]{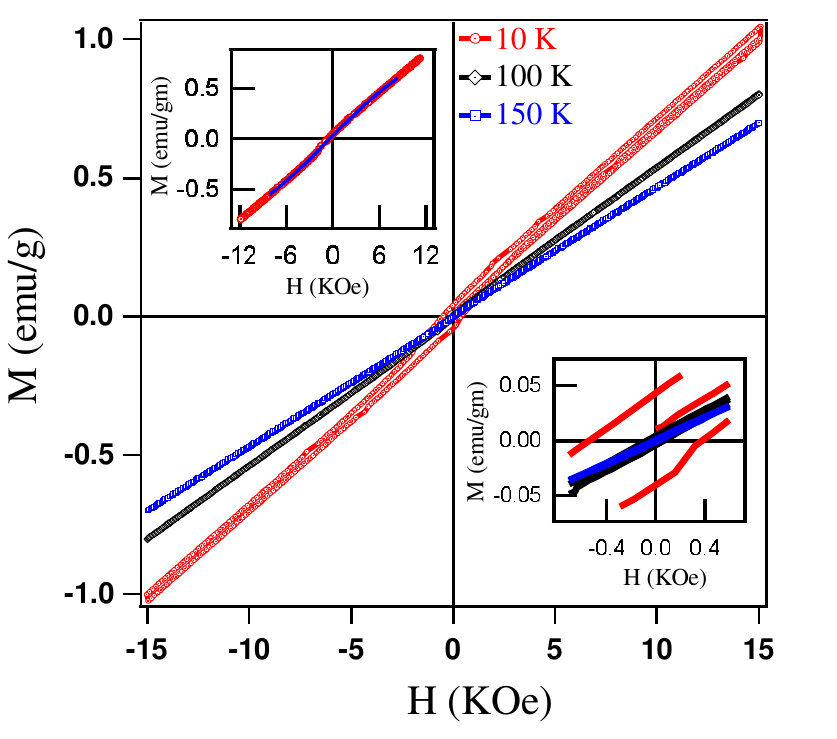}
  \caption{ Magnetization (M) as a function of applied magnetic field (H) is plotted for the ZFC mode at sample temperatures of 10, 100 and 150 K. Top-left inset shows M-H curve fit using the Eq.~\ref{eq2}. Bottom-right inset shows enlarged data around the zero magnetic field to show the hysteresis of M-H curve for 10 K, which disappears above 100 K.}
  \label{5}
\end{figure}

Figure~\ref{3} shows temperature dependent electrical resistivity of Cr$_{0.79}$Se measured within the temperature range of 3.1 to 310 K. We observe from the resistivity data that at low temperatures (T$<$41 K), the data nicely fits to the Fermi liquid law of resistivity ($\propto$ $aT^2$). But beyond, 41 K the data follows a sublinear behaviour ($\propto$ $bT^{0.62}$) with temperature up to 200 K. Inset, in Fig.~\ref{3} confirms the Fermi-liquid nature of the resistivity as one can notice perfect linear relation between $\rho$ and $T^2$ (for T up to 41K). Bottom panel of Fig.~\ref{3} presents the plot of d$\rho$/dT $vs$ T. We notice that d$\rho$/dT increases with T up to  41 K,  above 41 K d$\rho$/dT decreases with T up to 302 K, and beyond 302 K -d$\rho$/dT decreases with T hinting at an electronic phase transition at this temperature as d$\rho$/dT becomes $-ve$.  Next, in Figure~\ref{4}, we show magnetization (M) as a function of temperature measured under zero-field-cooled (ZFC) and field-cooled (FC) modes at an applied external magnetic field of 500 Oe. In Fig.~\ref{4}, further we show inverse magnetic susceptibility (1/$\chi$) as a function of temperature measured in the FC mode. As can be seen from Fig.~\ref{4}, at higher temperatures (T$>$225 K), susceptibility follows the Curie-Weiss law,

\begin{equation}\label{eq1}
 \chi (T)=\frac{C}{T-\Theta}
\end{equation}

here, $C$ is the Curie constant and $\Theta$ is the Curie-Weiss temperature. From the fitting, we found a Curie-Weiss temperature of $\Theta$=-300$\pm$2 K and a Curie constant of $C$= 24.5$\pm$0.5 Oe.gm.emu$^{-1}$.K$^{-1}$. The negative Curie-Weiss temperature suggest for a dominant antiferromagnetic interactions in the system.   We further have calculated the effective magnetic moment of Cr ion in the paramagnetic regime using the formula,  $\mu_{eff}=2.84 \sqrt{MC}$ $\mu_B=5.08\mu_B$/Cr~\cite{Makovetskii1978}.  In addition, we observe a deviation from the linear dependence of 1/$\chi$ on T below 225 K,  suggesting a magnetic transition from a paramagnetic to an antiferromagnetic phase.

\begin{figure*}
\centering
  \includegraphics[width=0.8\textwidth]{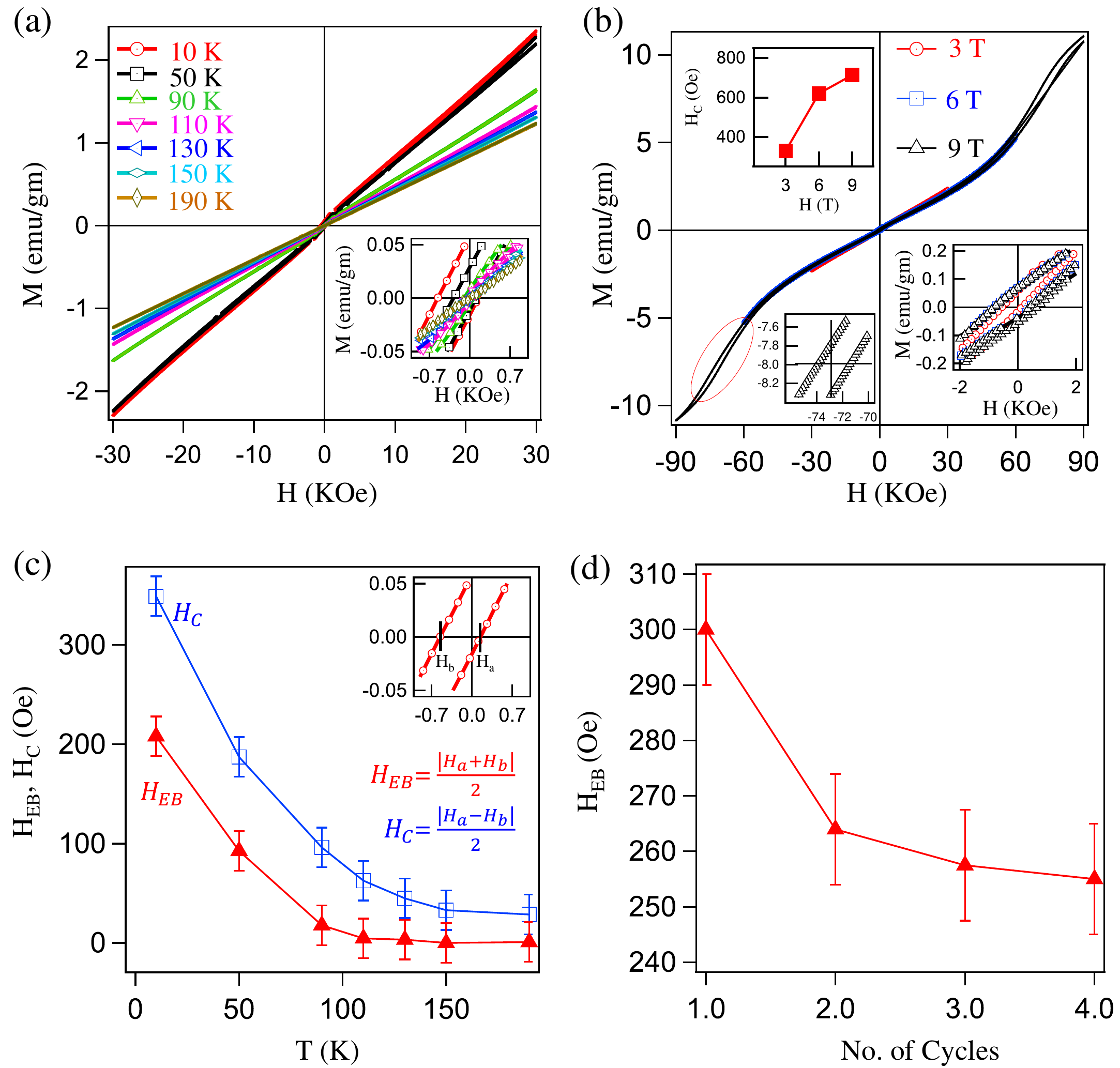}
  \caption{(a) M-H loops plotted for the FC mode at various sample temperatures. Inset in (a) is enlarged data around the zero magnetic field to show the hysteresis of M-H curve. (b) M-H curves plotted for the FC mode at various applied magnetic fields. Bottom-right inset in (b) is enlarged data around the zero magnetic field to show the hysteresis of M-H curve at various applied fields. Bottom-left inset in (b) is enlarged data to show the hysteresis of M-H curve for the applied field of 9 T. Top-left inset in (b) shows coercive field as a function of applied field.  (c) Exchange bias (H$_{EB}$) and coercive field (H$_C$) are plotted as a function of temperature measured with 3T magnetic field in the FC mode.  (d) Plot of H$_{EB}$ $vs.$ no. of M-H loop cycles.}
  \label{6}
\end{figure*}

Figure~\ref{5} depicts $M-H$ curves taken in the ZFC mode with an applied magnetic field of 1.5 T at temperatures 10, 100, and 150 K. From the inset shown at the bottom of Fig.~\ref{5}, we observe hysteresis in the $M-H$ loop with a coercive field of $H_C$=410 Oe when measured at 10 K and the hysteresis disappears at 100 K. Presence of hysteresis loop suggests for a ferromagnetic order at low temperature. Also, magnetization saturation is not reached up to the applied field of 1.5 T suggesting for a strong AFM order as well in this system. In order to quantify the strength of ferromagnetism in the system, as shown in the top-left inset of Fig.~\ref{5}, we performed M-H curve fitting with equation~\ref{eq2}~\cite{Patel2018}, to estimate the saturation magnetization M$_s$=1.3$\pm$0.1 emu/g and remanent magnetization  M$_r$=0.09$\pm$0.01 emu/g, while holding the experimental coercive field H$_c$=410 Oe and susceptibility $\chi$=9.1 $\times$ 10$^{-5}$ emu/(Oe-g). These values suggest for a weak ferromagnetism in Cr$_{0.79}$Se.

\begin{equation}\label{eq2}
M(H)= \frac{2M_s}{\pi} tan^{-1} [(\frac{H}{H_c}\pm1)~tan(\frac{\pi M_r}{2M_s})]+\chi H
\end{equation}

Figure~\ref{6}(a) depicts $M-H$ curves measured in the FC mode at various sample temperatures using the applied magnetic field of 3 T. Inset at the right-bottom  of Fig.~\ref{6}(a) demonstrate a significant shift in the M-H loop hysteresis, hinting at the presence of exchange bias in this sample. Figure~\ref{6}(b) depicts M-H curves in the FC mode at a fixed sample temperature of 10 K by varying the applied magnetic fields, 3 T, 6 T, and 9 T. Right-bottom inset of Fig.~\ref{6}(b) is the zoomed in image of $M-H$ loops in which one can notice hysteresis at all applied magnetic fields. Further, left-bottom inset of Fig.~\ref{6}(b) is the zoomed in image of $M-H$ loop measured with a magnetic field of 9 T,  demonstrating the presence of hysteresis between 6 and 9 T. Figure~\ref{6}(c) depicts the exchange bias ($H_{EB}$)  and coercivity ($H_C$) plotted as a function of temperature.  From Fig.~\ref{6}(c) we can observe that both $H_{EB}$ and $H_{C}$ decrease with increase in temperature and become negligible above 100 K within the instrumental error of $\pm$20 Oe. Figure~\ref{6}(d) depicts training effect on the exchange bias. We observe a significant decrease (15$\%$) in the exchange bias after repeating four cycles of $M-H$ loops. The observation of exchange bias under the FC mode but not under the ZFC mode is inline with the phenomena of the exchange bias, as explained for various magnetic alloys and compounds~\cite{Giri2011}.

\section{Discussions}

The studied sample,  Cr$_{0.79}$Se, has been found in the NiAs-type structure of the hexagonal crystal symmetry as demonstrated in Figs.~\ref{1} and ~\ref{2}, which is consistent with the crystal structure of  stoichiometric CrSe~\cite{Tsubokawa1956}. On the other hand, the crystal symmetry of the so far existing off-stoichiometric systems of Cr$_x$Se deviate form the stoichiometric system. For instance, Cr$_{0.67}$Se posses trigonal crystal symmetry with a space group of R$\bar{3}$H (148)~\cite{Adachi1994}, while Cr$_{0.875}$Se, Cr$_{0.75}$Se and Cr$_{0.625}$Se posses the monoclinic crystal symmetry with a space group of 	C1$_2$/m1(12)~\cite{Chevreton1961, Blachnik1987, Sleight1969}.  Most importantly,  we identify that Cr$_{0.79}$Se is the first known off-stoichiometric composition crystallizing in the P6$_3$/mmc space group. Further from the temperature dependent XRD data, we notice peak shifting with the temperature.  Such a peak position shift with temperature generally lead to change in the lattice parameters while still preserving the crystal symmetry. This has been supported by the rietveld refinement (see Fig.~\ref{2}). Thus, our studies confirm that the NiAs-crystal structure of Cr$_{0.79}$Se is stable up to 600$^o$C from the room temperature. Though Jahn-Teller distortion is not observed so far in any of the off-stoichiometric Cr$_x$Se compositions, in the case of stoichiometric CrSe, the Jahn-Teller distortion is suggested for temperature below 305$^o$C~\cite{Masumoto1962}. However, we do not observe any signature of Jahn-Teller distortion from our temperature dependant XRD data in the studied compound.

Next, from the electrical resistivity data shown in Fig.~\ref{3}(a) it is clear that Cr$_{0.79}$Se is as a Fermi-liquid type metal below 41 K. But above 41 K, the system deviates to a non-Fermi-liquid type metal showing a sub-linear dependence of the resistivity on temperature up to 200 K.  The same is confirmed from the $d\rho/dT$ $vs$ T curve [see bottom of Fig.~\ref{3}(a)]. From this curve we observe that $d\rho/dT$ increases with temperature up to 41 K and after that it decreases with $T$ to reach zero at 302 K.  Above 302 K, -$d\rho/dT$ decreases with increasing $T$ up to the measured temperature of 310 K. This observation hints at a metal-insulator (MI) transition above 302 K in Cr$_{0.79}$Se. MI transition has been noticed in some of the other transition-metal monochalcogenides as well. For instance, in Fe$_{0.875}$Se the MI transition is observed at 100 K due to a proximity effect of magnetic moment reorientation~\cite{Li2016}. Note here that a similar kind of resistivity data have been reported earlier in the case of Cr$_{0.67}$Se single crystals~\cite{Wu2020} with a sub-linear behaviour of resistivity up to 175 K and then changing the resistivity slop above this temperature. But note that the antiferromagnetic ordering in Cr$_{0.67}$Se is found only below 60 K, while in our sample the antiferromagnetic ordering is found at 225 K. Another report on the Cr$_{0.68}$Se single crystals also suggested for an antiferromagnetic ordering below 42 K,  but unlike to Cr$_{0.67}$Se which is a metal up to 300 K, Cr$_{0.68}$Se is found to be a small gap semiconductor ($E_g$=3.9 meV)~\cite{Yan2017} down to lowest possible measured temperature. Thus, the electrical properties of Cr$_x$Se systems seem to be highly sensitive to the Cr atom concentration,  which may directly affect the charge carrier density near the Fermi level and disorder of the system,  rather than the magnetic interactions as we do not find any one-one correlation between the magnetic transition temperatures (see Figs.~\ref{4} -~\ref{6}) and the temperature dependent resistivity (see Fig.~\ref{3}).

Finally  coming to the important observations of this study, we found a weak ferromagnetism in Cr$_{0.79}$Se  below T$_C$=100 K that is coexisted with the AFM phase. As a result an exchange bias has been observed in this system below T$_C$.  Usually, CrSe is known for their non-collinear AFM phase. But recently, one report showed a weak ferromagnetism in  Cr$_{0.67}$Se along with AFM phase below 50 K~\cite{Wu2020}. However not much discussion is drawn on the presence of the exchange bias in  Cr$_{0.67}$Se. Thus, we  report the exchange bias for the first time in these systems.  Another report on the neutron diffraction studies of Cr$_{0.67}$Se suggests for two magnetic phases, non-collinear AFM phase at low temperature ($<$ 38 K) and collinear AFM phase at high temperature (38 K$>$T$<$45 K), but did not find the ferromagnetism down to as low as 6 K~\cite{Adachi1994}. On the other hand, the estimated effective paramagnetic moment of 5.08$\mu_B$/Cr in our systems is slightly higher than the effective paramagnetic moment of 4.5$\mu_B$/Cr in the stoichiometric CrSe~\cite{Lotgering1957}. This could be mostly because of the mixed Cr valance states in the off-stoichiometric compositions ~\cite{Andresen1970, Liu2017}. Although, earlier a spin-glass like magnetic phase has been observed due to the frustrated magnetic moments between AFM and FM phases~\cite{Li2006}, in our system from the $M-T$ data (see Fig.~\ref{4}) we do not see any signature of the spin-glass like behaviour below T$_C$.

\section{Conclusions}
In summary,  we systematically studied the structural, electrical transport, and magnetic properties of the antiferromagnetic transition-metal monochalcogenide Cr$_{0.79}$Se. We identify that Cr$_{0.79}$Se is synthesised into the same NiAs-type hexagonal crystal structure of the stoichiometric CrSe, unlike the other off-stoichiometric systems which form in differing crystal symmetries. Resistivity data suggest Cr$_{0.79}$Se to be a Fermi-liquid-type metal at low temperatures, while at higher temperatures the resistivity depends sublinearly on the temperature.  Above the room temperature the resistivity data hints at a MI transition, but need more studies to confirm the same.  Magnetic measurements suggest for a transition from paramagnetic phase to an antiferromagnetic phase at a N$\acute{e}$el temperature of 225 K. Importantly, a weak ferromagnetism is noticed below 100 K along with the antiferromagnetism.  As a result, we notice significant exchange bias below 100 K due to the interaction between the ferro- and antiferromagnetic phases.

\section{Methods}

 Samples of Cr$_{0.79}$Se are prepared by the standard solid-state reaction method~\cite{Tsubokawa1956} from high purity powders of Chromium (4N, Alfa Aeser) and Selenium (5N, Alfa Aeser) elements by mixing in appropriate ratio. The well-mixed powders were then heated in a muffle furnace at 1000$^o$C for 48 hours. The final sample was pressed into pellet form and heated again at 1000$^o$C for another 48 hours.  As prepared polycrystalline sample was structurally characterized using the powder X-ray diffraction (XRD) equipped with Cu K$\alpha$ radiation of Rigaku-SmartLab (9 KW) at various samples temperatures (30$^o$C to 600$^o$C). Rietveld refinement analysis of the XRD data is done using FULLPROF software package~\cite{RodriguezCarvajal1993}. Energy-dispersive X-ray (EDX) analysis suggests the chemical composition of as prepared sample to be Cr$_{0.79}$Se. Electrical resistivity measurements were carried out using the standard four-probe technique with a closed-cycle refrigerator (CCR) based cryostat, within a temperature range of 3.1 K to 310 K. Conducting silver epoxy and Cu wires were used to make the electrical contacts. Magnetic property measurements were carried out using the vibrating sample magnetometer (VSM) (DynaCool, Quantum Design) up to a magnetic field of 9 tesla.

\section{Acknowledgements}
S.T. acknowledges the financial support given by SNBNCBS through the Faculty Seed Grants program. Authors thank Science and Engineering Research Board (SERB), Department of Science and Technology (DST), India  for the financial support through the start-up research grants (SRG/2020/000393).

\section{Conflicts of Interest}
The authors declare no conflicts of interest.

\bibliographystyle{achemso}
\bibliography{CrSe}

\begin{figure*}
\centering
  \includegraphics[width=0.8\textwidth]{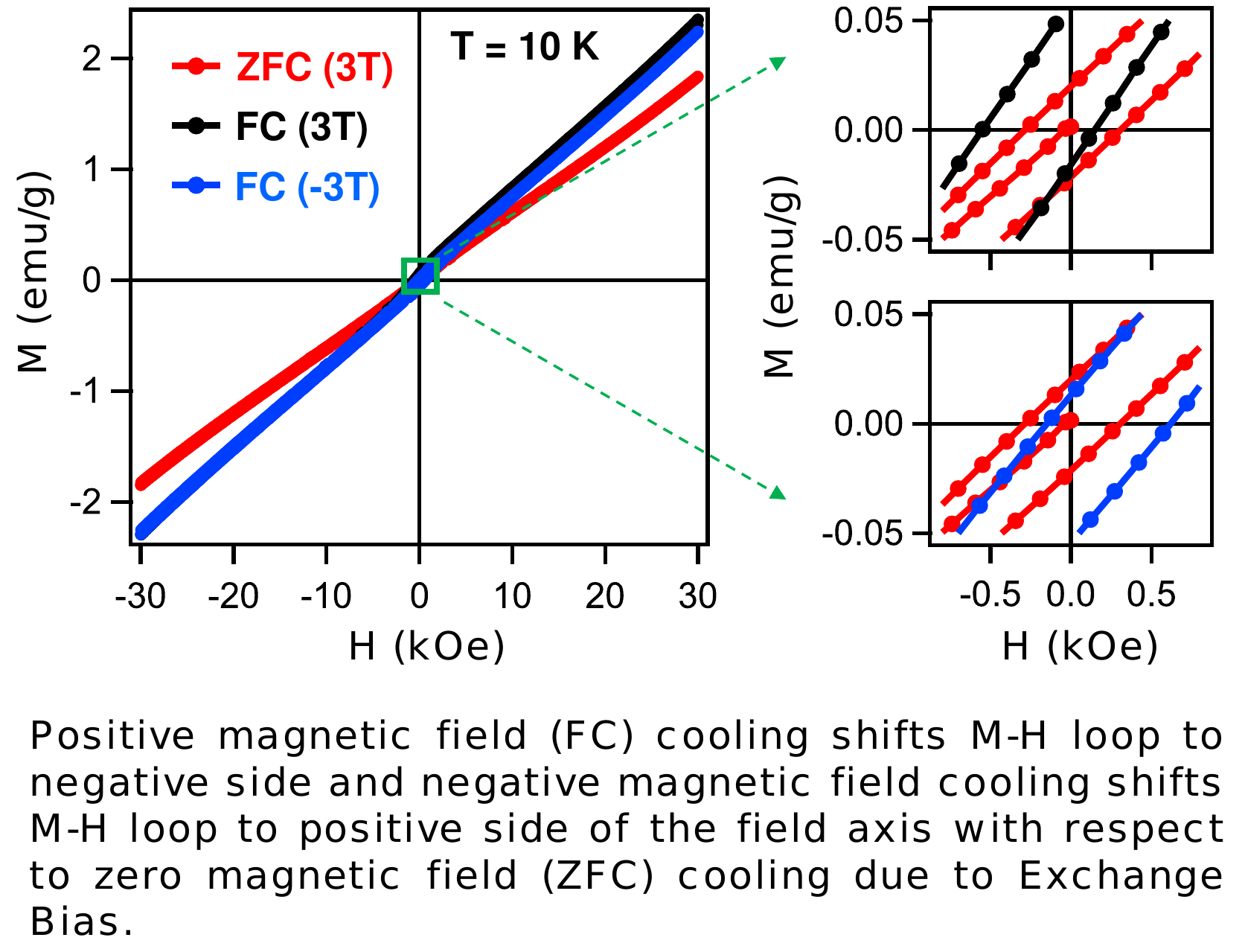}

\end{figure*}

\end{document}